\documentclass[aps,prl,twocolumn,showpacs,floatfix]{revtex4-1}
\usepackage[utf8]{inputenc}
\usepackage[T1]{fontenc}
\usepackage{float} 
\usepackage{color}  
\usepackage{graphicx}
\usepackage{amsmath}
\usepackage{amssymb}
\usepackage{bbm}
\usepackage{indentfirst}
\usepackage{dcolumn}
\usepackage{soul}
\usepackage{mathrsfs}
\usepackage{xcolor}
\usepackage{ulem}

\usepackage{relsize}
\usepackage{amsmath}
\usepackage{graphicx}
\usepackage{float}

\usepackage{color}
\makeatletter

\usepackage{xcolor}
\usepackage{soul}
\usepackage{soulpos}

\usepackage{dcolumn}
\usepackage{bm}
\usepackage{subfigure}
\usepackage{helvet}
 \usepackage{color}
\graphicspath{{figures/}}

\begin{document}

\title{Universal criteria for single femtosecond pulse ultrafast magnetization switching in ferrimagnets}

 \author{F. Jakobs}
\affiliation{Dahlem Center for Complex Quantum Systems and Fachbereich Physik, Freie Universit\"{a}t Berlin, 14195 Berlin, Germany}
\author{U. Atxitia}
\email{unai.atxitia@fu-berlin.de.}
\affiliation{Dahlem Center for Complex Quantum Systems and Fachbereich Physik, Freie Universit\"{a}t Berlin, 14195 Berlin, Germany}

\begin{abstract}
Single-pulse switching has been experimentally demonstrated in ferrimagnetic GdFeCo and  Mn$_2$Ru$_x$Ga alloys.
Complete understanding of single-pulse switching is missing due to the lack of an established theory accurately describing the transition to the non-equilibrium reversal path induced by femtosecond laser photo-excitation.
In this work we present general macroscopic theory for the magnetization dynamics of ferrimagnetic materials upon femtosecond laser excitation. 
Our theory reproduces quantitatively all stages of the switching process observed in experiments.
We directly compare our theory to computer simulations using atomistic spin dynamics methods for both GdFeCo and Mn$_2$Ru$_x$Ga alloys.  
We provide explicit expressions for the magnetization relaxation rates in terms of microscopic parameters which allows us to propose universal criteria for  switching in ferrimagnets.
\end{abstract}

\maketitle

\normalem

Ultrafast magnetization switching induced by a single femtosecond laser pulse has attracted a lot of attention as a promising solution for low energy, faster memory applications.~\cite{StanciuPRL2007,RaduNature2011,OstlerNatComm2012,LeGuyader2012,GravesNatMaterials2013,LiuNanoLetters2015,Hadri2016,Lalieu2017,Chen2017,Lalieu2019,IacoccaNatComm2019,Kimel2019,Ignatyeva2019,VanHees2020,Davies2020}. Until recently only GdFeCo alloys and synthetic ferrimagnets \cite{Lalieu2017}, presented the ability to switch under  either optical femtosecond laser- \cite{OstlerNatComm2012,LeGuyader2012,Lalieu2017} or electric picosecond current-pulses \cite{Yang2017,El-Ghazaly2020}. Although several micro- and macroscopic models have reproduced single-pulse switching in GdFeCo ferrimagnets \cite{Vahaplar2009,MentinkPRL2012,Barker2013,WienholdtPRB2013,Schellekens2013,Baryakhtar2013,Baral2015,Gridnev2016,Kalashnikova2016,Zhang2016,Krivoruchko2016,Gerlach2017,Gridnev2018,Beens2019,Vogler2019,Davies2020b}, complete understanding of the role of electrons, lattice and spin sublattices and their mutual interactions remains a challenge \cite{Carva2017}. The existing criteria for switching rely on the existence of two antiferromagnetically coupled magnetic sublattices showing distinct dynamical response to femtosecond laser photo-excitation. While in single species ferromagnets such as 3d transition metals, relaxation of angular momentum occurs via dissipation into other degrees of freedom -- relativistic relaxation -- in two-sublattice magnets, additionally, relaxation can occur via angular momentum exchange between sublattices -- exchange relaxation. By driving the spin system into a non-equilibrium state where exchange relaxation dominates, a non-equilibrium ultrafast reversal path opens. One of the most outstanding, open questions is about the conditions or criteria for the onset of the exchange dominated relaxation regime. Crucially, it is unclear how previous understanding gained from observation in GdFeCo translates into the recent discovery of single-pulse switching in the ferrimagnetic Heusler alloy Mn$_2$Ru$_x$Ga \cite{Banerjee2020}, where the two antiferromagnetically coupled Mn atoms are of the same kind in comparison to Gd and FeCo. 

In this work we present a general theoretical framework for the description of single pulse switching of ferrimagnets. We provide explicit expressions for the relativistic and exchange relaxation parameters  as a function of microscopic material parameters, including their dependence on temperature and non-equilibrium sublattice magnetization. This allows us to uncover the criteria for the onset of the exchange-dominated relaxation regime and switching. We verify the validity of the model by direct comparison to atomistic spin model simulations of both GdFeCo and Mn$_2$Ru$_x$Ga alloys. 

We shall describe each magnetic atom at site $i$ as a classical spin vector $\mathbf{s}$ of a unit length. The magnetic and mechanical moments of each atom/element are given by $\boldsymbol{\mu} = \mu_{s} \mathbf{s}$ and $\mathbf{S}=\mu_{s} \mathbf{s}/\gamma$ where $\mu_s$ is the magnetic moment and $\gamma$ is the gyromagnetic ratio. We consider a classical Heisenberg spin Hamiltonian\cite{Nowak2007BOOK}: 
%============ 
\begin{equation}   
\mathcal{H}= - \sum_{i \neq j \\
\langle ij \rangle} J_{ij} \mathbf{s}_i \cdot \mathbf{s}_j - \sum_{i} d^z_i (s^z_i)^2.
\label{eq:Ham}
\end{equation}
%=========== 
To model a ferrimagnet, one needs to consider two sublattices  with different and antiparallel magnetic moments $\mu_a$ and $\mu_b$, with three different exchange coupling constants. Two ferromagnetic couplings for each sublattice coupling with itself ($J_a$ and $J_b$ >0) and a third for the antiferromagnetic interaction between them, $J_{ab}<0$ \cite{OstlerPRB2011}. The second term in Eq. \eqref{eq:Ham} describes the contribution to the energy of on-site uniaxial anisotropy with easy-axis in $z$-direction and anisotropy constant, $d^z_i$.

The macroscopic model we propose is derived from a microscopic spin model, where the equation of motion for the spin dynamics of each atomic spin is the stochastic Landau-Lifshitz-Gilbert (LLG) equation~\cite{Nowak2007BOOK}: 
%============ 
\begin{equation}
\frac{\partial \mathbf{s}_i}{\partial t} = - \frac{|\gamma|}{(1+\lambda^2)\mu_{i}}\left[\left( \mathbf{s}_i \times \mathbf{H}_i \right) - \lambda  \left( \mathbf{s}_i \times \left(\mathbf{s}_i \times \mathbf{H}_i \right) \right)\right].
\label{eq:llg}
\end{equation}
%=========== 
$\lambda$ is the local intrinsic atomic damping, the effective field $\mathbf{H}_i= \boldsymbol{\zeta}_i - \frac{\partial \mathcal{H}}{\partial \mathbf{s}_i}$, where  thermal fluctuations are represented by the stochastic field $\boldsymbol{\zeta}_i$. 

\begin{figure}[!tb]
\includegraphics[width=8.6cm]{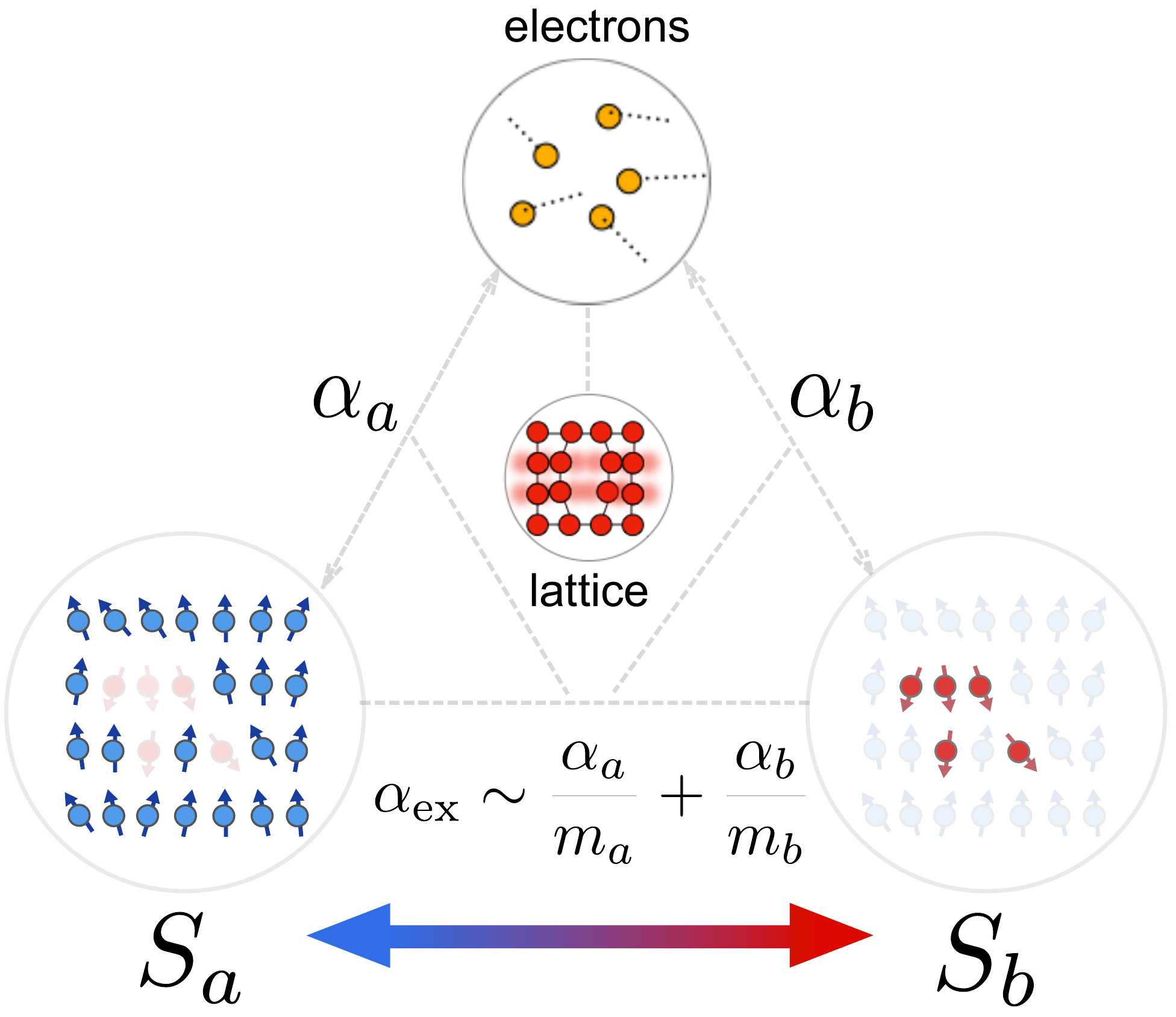}
\caption{Degrees of freedom involved in single-pulse switching; electrons, lattice and element-specific spin systems. Electrons act as heat-bath to the spin system with an element-specific coupling strength given by $\alpha_{a(b)}$. Exchange relaxation rate $\alpha_{\rm{ex}}\sim \alpha_a/m_a+\alpha_b/m_b$, where  $m_{a(b)}$ are the normalized sublattice magnetization magnitude.  
}
\label{fig:sketch}
\end{figure}

The non-equilibrium macroscopic magnetization dynamics of the element-specific angular momentum $S_a = \mu_a \langle s_a \rangle /\gamma$, where $m_a=\langle s_a \rangle$ is the first moment of the non-equilibrium distribution function, can be described by \cite{MentinkPRL2012}:
%=====================
\begin{equation}
\frac{d S_a}{d t} =  \alpha_{a} \mu_a  H_a + \alpha_{\rm{ex}} ( \mu_a H_a - \mu_b H_b )
\label{eq:long-LLB-1-ex}
\end{equation}
%=====================
where $a \neq b$. The macroscopic relativistic damping parameter in Eq. \eqref{eq:long-LLB-1-ex} reads \cite{Atxitia2012}
%=====================
\begin{equation}
\alpha_{a} = 2\lambda_a \frac{L(\xi_a)}{\xi_a}.
\label{eq:alpha-a}
\end{equation}
%=====================
Here, $L(\xi)$ stands for the Langevin function. The relaxation parameter strongly depends on temperature and non-equilibrium magnetization state through the thermal field $\xi_a= \beta \mu_a H_a^{\rm{MFA}}$. In the exchange approximation, the MFA field acting on sublattice $a$ is:
%=======================
\begin{equation}
\mu_a H_a^{\rm{MFA}} = z_a J_{aa} m_a + z_{ab} J_{ab} m_b.
\label{eq:MFA-LLB}
\end{equation}
%======================
 Here, $z_a$ is the number of nearest neighbours (n.n.) of spins of type $a$, and $z_{ab}$ is the amount of n.n. of type $b$. The macroscopic damping increases with temperature up to a value $\alpha_a = 2/3$ at the critical temperature \cite{Atxitia2012}. The exchange relaxation parameter $\alpha_{\rm{ex}}$ is given by
%=====================
\begin{equation}
\alpha_{\rm{ex}} = \frac{1}{2}\left(\frac{\alpha_{a}}{z_{ab} m_a} +  \frac{\alpha_{b}}{z_{ab} m_b} \right).
\label{eq:alpha-ex}
\end{equation}
%=====================
This expression is the extension of the non-local exchange relaxation in ferromagnets to local exchange relaxation in ferrimagnets. The role of $\alpha_{\rm{ex}}$, $\alpha_a$ and $\alpha_b$ as the coupling between the sublattices and heat baths is visualized in Fig.~\ref{fig:sketch}. In  single species ferromagnets, sublattices $a$ and $b$ represent the same spin lattice, hence $\alpha_a=\alpha_b$. Therefore, $\alpha_{\rm{ex}}= \alpha_a/(z m_a)$, and $\mu_aH_a - \mu_bH_b = \mu_a H_{\rm{ex}} a_0^2 \Delta m_a$, with $a_0$ representing the lattice constant. Hence, $\Gamma_{\rm{ex}}^{\rm{non-loc.}}=\alpha_{\rm{ex}} (\mu_a H_a - \mu_b H_b) = \alpha_a (A/M_a(T)) \Delta m_a$, where $A$ is the so-called micromagnetic exchange stiffness \cite{Atxitia2010}.  $M_a(T)=(\mu_a/\upsilon_a)m_a$ is the magnetization density at temperature $T$, where $\upsilon_a$ is the unit cell volume. Non-local exchange relaxation plays a minimal role in the field of ultrafast magnetization dynamics since $\Delta m_a \ll 1$.

The non-equilibrium fields ($\mu_a H_a = 0$ at equilibrium) are given by 
%=====================
\begin{equation}
\mu_a H_a = \frac{(m_a-L(\xi_a))}{\beta L'(\xi_a)}.
\label{eq:muaHa}
\end{equation}
%=====================
where, $L'(\xi) = dL/d\xi$. Note that they are different to the MFA fields in Eq. \eqref{eq:MFA-LLB}, and have been derived previously \cite{Garanin1997,Atxitia2012}. As the magnetic system approaches thermal equilibrium, the non-equilibrium fields can be cast into Landau-like expressions \cite{MentinkPRL2012,Atxitia2012}. 

%%%%%%%%%%%%%%%%%%%%%%%%%%%%%%%%%%%%%%%%%%%%%%
\begin{figure}[!t]
\includegraphics[width=7.2cm]{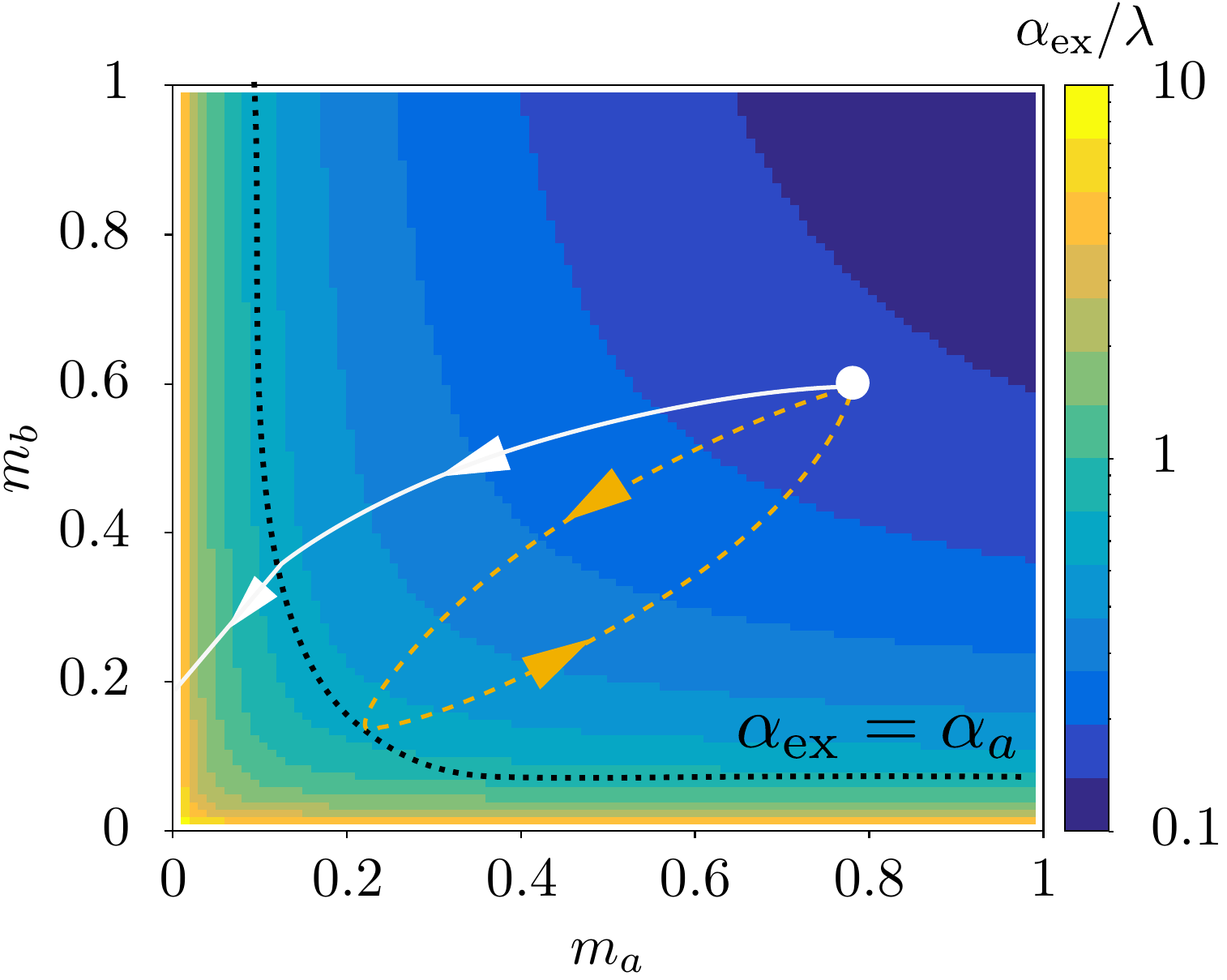}
\caption{Normalized exchange relaxation parameter  $\alpha_{\rm{ex}}/\lambda$ as function of the sublattice magnetization $m_a$ and $m_b$, at $T=600$~K. System parameters correspond to GdFeCo. $\lambda_a=\lambda_b=\lambda=0.01$. The dotted black line corresponds to  $\alpha_{\rm{ex}}=\alpha_a$. The white dot represents the starting sublattice magnetization ($m_a,m_b$). The closed dashed orange line describes a trajectory meeting a no-switching criteria. The open solid white line describes a trajectory meeting a switching criteria.}
\label{fig:alphaexdiagram}
\end{figure}
%%%%%%%%%%%%%%%%%%%%%%%%%%%%%%%%%%%%%%%%%%%%%%

Equation \eqref{eq:long-LLB-1-ex} has been proposed before based on symmetry arguments \cite{MentinkPRL2012,Baryakhtar2013} as a direct generalization of the Landau-Lifshitz equation with longitudinal relaxation terms. These models have introduced the relaxation parameters at a purely phenomenological level and to some extent their values are arbitrary. Moreover, since they were taken as constant values, most of the non-equilibrium spin physics was not taken into account. Our model overcomes these assumptions by providing expressions for the relativistic and exchange relaxation parameters as a function of the sublattice specific atomic relaxation parameter, $\lambda_{a(b)}$, and normalized magnetization $m_{a(b)}$. 
%%%%%%%%%%%%%%%%%%%%%%%%%%%%%%%%%%%%%%%%%%%
 \begin{figure*}[!tb]
\includegraphics[width=17.2cm]{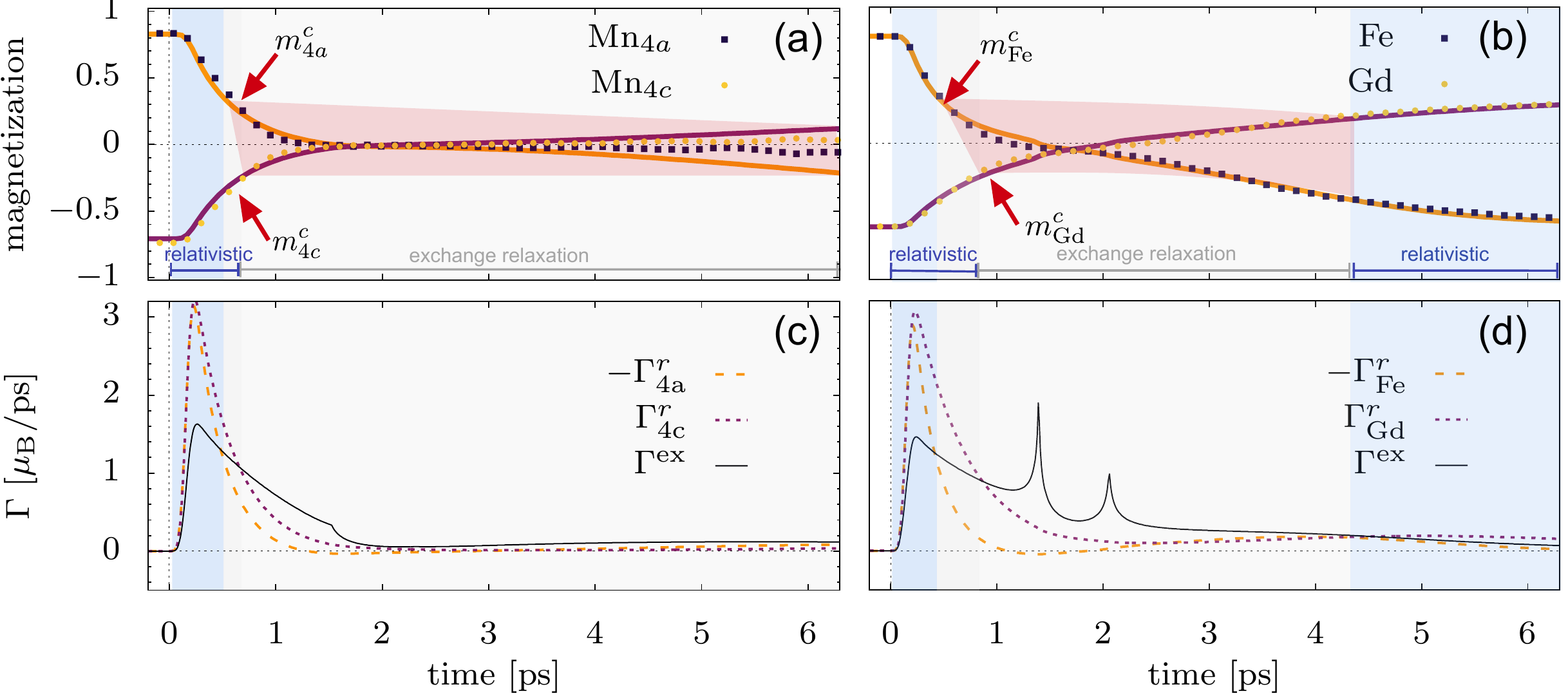}
\caption{Element-specific magnetization dynamics of single-pulse switching in ferrimagnets using an atomistic spin dynamics model (symbols) and a macroscopic model (solid lines) for Mn$_2$Ru$_{0.86}$Ga (a) and Gd$_{25}$FeCo (b). The red area corresponds to the critical values ($m_a^c,m_b^c$) to enter exchange relaxation regime. The blue area corresponds to the time lapse where relativistic relaxation dominates. Different contributions to the element-specific magnetization relaxation for the single-pulse switching dynamics in Mn$_2$Ru$_{0.86}$Ga (c) and Gd$_{25}$FeCo (d). $\Gamma^r_{a(b)}$ stands for the relativistic relaxation rate of sublattice $a=4a$ (c) and $a=$Fe (d), and $b=4c$ (c) and $b=$Gd (d). $\Gamma^{\rm{ex}}$ is the exchange relaxation rate, which is the same for both sublattices. $\lambda_a=\lambda_b=\lambda=0.01$.}
\label{fig:ASD-model-simulations}
\end{figure*}
%%%%%%%%%%%%%%%%%%%%%%%%%%%%%%%%%%%%%%%%%%%

This insight is paramount to find the criteria for the onset of a exchange relaxation dominated state. In Fig. \ref{fig:alphaexdiagram} we present a diagram of the normalized exchange relaxation parameter $\alpha_{\rm{ex}}/\lambda$ for GdFeCo alloy parameters as a function of $m_a$ and $m_b$ at a fixed temperature, $T=600$ K, which corresponds to the Curie temperature of the alloy. We observe that the exchange relaxation parameter strongly depends on the magnitude of sub-lattice magnetization and its value ranges over three orders of magnitude, $\alpha_{\rm{ex}}/\lambda= 0.1 - 10$. Importantly, only when magnetic states reduce significantly, does the exchange relaxation dominates over relativistic relaxation. 
 
So far only two classes of ferrimagnets have shown single-pulse switching, Gd$_x$(FeCo)$_{1-x}$ alloys and Mn$_2$Ru$_x$Ga. Switching in GdFeCo has been thoroughly studied both experimentally~\cite{StanciuPRL2007,RaduNature2011,OstlerNatComm2012,LeGuyader2012,GravesNatMaterials2013} and theoretically~\cite{OstlerNatComm2012,Barker2013,Schellekens2013,WienholdtPRB2013,Baryakhtar2013,Beens2019,IacoccaNatComm2019,Vogler2019}, whereas in Mn$_2$Ru$_x$Ga has been only recently demonstrated for a range of Ru concentrations ($x\geq 0.9$) \cite{Banerjee2020}. In GdFeCo alloy, switching is characterised by a fast response of the Fe sublattice and slower response of the Gd sublattice to  femtosecond laser photo-excitation. This difference roots to their distinct magnetic moment, $\mu_{\rm{Gd}}=7.6 \mu_{\rm{B}}$ and $\mu_{\rm{Fe}}=1.92 \mu_{\rm{B}}$. Differently to GdFeCo, antiferromagnetically coupled Mn spins in Mn$_2$Ru$_x$Ga have similar atomic magnetic moments. We demonstrate the universality of our theory by direct comparison of the photo-induced magnetization switching to computer simulations based on atomistic spin dynamics (Eq. \eqref{eq:llg}) for GdFeCo (disordered spin structure) and Mn$_2$Ru$_x$Ga (ordered spin structure). Their magnetic properties such as magnetic moments and exchange interactions differ substantially. We consider two typical  compositions, that show switching, Gd$_{25}$(FeCo)$_{75}$ alloys and Mn$_2$Ru$_{0.86}$Ga. It is noteworthy; that while for GdFeCo atomistic spin models have been used thoroughly, in Mn$_2$Ru$_{0.86}$Ga similar simulations are missing. We derive the necessary material parameters based on experimental measurements~\cite{Fowley2018,Betto2015,Thiyagarajah2015, Kurt2014, Bonfiglio2021}. We find that for Mn$_2$Ru$_{0.86}$Ga the atomic magnetic moments, $\mu_{4a}$ = 2.88 $\mu_{\rm{B}}$ and $\mu_{4c}=$ 4.05 $\mu_{\rm{B}}$, and the exchange parameters, $J_a=1.28 \times 10^{21}$ J, $J_b=4.0\times 10^{22}$ J and $J_{ab}=-4.85\times 10^{22}$ J describe the equilibrium magnetization well as function of temperature. Using these parameters the Curie temperature becomes $T_c=600$~K and the compensation temperature $T_{\rm{M}} \approx 300$~K. We note that in the experimental samples, those temperatures are sensitive to the growth conditions as well as material composition. However our temperatures are within the reported range of experimentally found temperatures~\cite{Fowley2018,Betto2015,Thiyagarajah2015, Kurt2014, Bonfiglio2021}. We use the so-called two-temperature model (TTM) to describe the dynamics of the electron and phonon systems\cite{Kaganov1957,Chen2006}. For both materials we use the same parameters in the TTM and thus the electron and phonon temperatures are the same for the same fluence in both materials \cite{Bonfiglio2021}. The heat-bath to which the spins are coupled is represented by the electron system. 

Figures \ref{fig:ASD-model-simulations} (a) and (b) show excellent agreement between our macroscopic model and  atomistic spin dynamics simulations for both alloys for all stages of the magnetization dynamics leading to switching, from fast demagnetization, transient ferromagnetic-like state, to magnetization recovery. Interestingly, the switching dynamics in Mn$_2$Ru$_{0.86}$Ga (Fig. \ref{fig:ASD-model-simulations} (a)) differs to GdFeCo (Fig. \ref{fig:ASD-model-simulations} (b)), both Mn sublattices demagnetize at similar rate and switch almost simultaneously. Although demagnetization timescales are similar in Mn$_2$Ru$_{0.86}$Ga and GdFeCo, the recovery of the magnetization in Mn$_2$Ru$_{0.86}$Ga is significantly slower. The relaxation of the sublattice magnetization (Eq. \ref{eq:long-LLB-1-ex}) can be split into two contributions, the relativistic relaxation: $\Gamma^r_a =\alpha_a\mu_a H_a$, and exchange relaxation $\Gamma^{\rm{ex}}=\alpha_{\rm{ex}}(\mu_a H_a-\mu_b H_b)$ (see Fig. \ref{fig:ASD-model-simulations}(c) and (d)). For both ferrimagnetic alloys, $\Gamma_a^r$ drives the dynamics in the first hundreds of femtoseconds, until sublattice magnetization reduces sufficiently to enter the exchange relaxation dominated region  $\Gamma^{\rm{ex}} > \Gamma_a^r$ (Fig. \ref{fig:alphaexdiagram}). In this regime, the exchange relaxation steers the systems towards an intermediate metastable state defined by the condition $\mu_a H_a=\mu_b H_b$. Under some conditions this intermediate state precede switching. It is cumbersome however to directly analyse Eqs. \eqref{eq:muaHa} due to its highly non-linear character. Therefore in the following we investigate some limiting scenarios.

In the high temperature limit, $\xi_a= \beta \mu_a H_a^{\rm{MFA}}\rightarrow 0$, $\Gamma^r_a = 2 \lambda_a k_BTm_a$, which is the so-called thermal fluctuation field. The corresponding relaxation time, $\tau_a = \mu_a/ (\gamma 2 \lambda_a k_BT)$, associated to relativistic relaxation, has been discussed before \cite{MentinkPRL2012,Radu2015}. Similarly, we can estimate the high-temperature limit of the exchange relaxation rate 
\begin{equation}
    \Gamma^{\rm{ex}}_{\infty} (m_a,m_b)= \lambda \frac{k_B T}{z} \frac{(m_a+m_b)(m^z_a-m_b^z)}{m_am_b}.
    \label{eq:Gamma-ex}
\end{equation}
High temperature limits are valid when the temperature is larger than the exchange energy acting on the spins. Otherwise, intermediate-to-high temperature limit are necessary. This limit adds corrections to previous estimations. For instance, $\mu_a H_a- \mu_b H_b = 3 k_{B} ((T-T_c^a) m_a+ (T-T_c^b) m_b)$, where $J_{0,a}+J_{0,ab}=3k_{B} T^a_c$. The exchange relaxation rate is
\begin{equation}
    \Gamma^{\rm{ex}}_{T} (m_a,m_b)= \Gamma^{\rm{ex}}_{\infty} \left(1 - \frac{1}{T}\frac{T_c^a m_a +T_c^bm_b}{m_a+m_b} \right)
    \label{eq:Gamma-ex-T}
\end{equation}
For element-specific critical temperature $T_c^a \approx T_c^b$, the correction can be cast as $\Gamma^{\rm{ex}}_{T}=\Gamma^{\rm{ex}}_{\infty}(T-T^a_c)/T$.Since the relativistic relaxation rate is also modified as $\Gamma^r_{a}=\Gamma^{r}_{a,\infty}(T-T^a_c)/T$, we can fairly investigate the crossover from relativistic to exchange dominated regime by comparing $\Gamma^{\rm{ex}}_{\infty}$ and $\Gamma^{r}_{a,\infty}$. Two cases of interest exist, i) one sublattice is faster than the other, and ii) both sublattices demagnetize at the same rate. 

In the first case, $\tau_a \ll \tau_b$, sublattice $a$ demagnetizes faster than sublattice $b$. This corresponds to GdFeCo (Fig. \eqref{fig:ASD-model-simulations}(b)). Soon after the application of a fs laser pulse, $m_a\ll m_b$, and consequently, $\Gamma^{\rm{ex}} \approx \lambda k_B T m_b/(z m_a)$ (cf. Eq.\eqref{eq:Gamma-ex}). We estimate the conditions for the transition from relativistic to exchange-dominated regimes (see Fig. \eqref{fig:ASD-model-simulations}(d)): From $\Gamma^{\rm{ex}} > \Gamma^r_{a}$,  $m_a < \sqrt{m_b/2z_{ab}} \approx 0.288 \sqrt{m_b}$, and from $\Gamma^{\rm{ex}} > \Gamma^r_{b}$ and $m_a \leq 1/2z =0.0833$ (red colored area in Fig. \ref{fig:ASD-model-simulations}~(b)). A second case, $\tau_a \approx \tau_b$, might also arise when demagnetization times of both sublattices are similar. This is the case of Mn$_2$Ru$_{0.86}$Ga alloys (Fig. \eqref{fig:ASD-model-simulations}(c)). One can estimate the conditions for the transition  $\Gamma^{\rm{ex}} = \Gamma_a^r$. This happens for both sublattices when $m_{a,b} = (m_a(0)+m_b(0))^2/(m_a(0)m_b(0))/2z_{ab}$. Assuming a realistic value of $m_b(0)/m_a(0)=0.9$, the exchange-dominate regime is reached when $m_{a,b}=0.334$ (red colored area in Fig. \ref{fig:ASD-model-simulations} (c)). Notably, this condition only depends on the initial values of the magnetization, $m_{a(b)}(0)$. These results are simple and general, and one of the main result of the present work. One can interpret the transition from relativistic to exchange-dominated regime as follows: as the magnetization of one sublattice decreases, the phase space of available states for the spins of the other sublattice to switch by exchange of angular momentum dramatically increases. While switching spin via coupling to an external bath becomes increasingly difficult as the magnetization reduces. 

Once the system has entered the exchange dominated regime the dynamics follows a path where total angular momentum is conserved  towards a magnetic state where  $\mu_a H_a = \mu_b H_b$. In the high temperature limit, this condition reduces to $m^z_a=m^z_b$, meaning that the exchange relaxation drives the magnetization of both sublattices into the same polarity, i.e. a ferromagnetic-like state \cite{RaduNature2011}. From this condition arises, that in order to have a final $m_a^z < 0$ the following condition is necessary: $S^{\rm{ex}}_a+S_b^{\rm{ex}}<0$. Here $S_{a(b)}^{\rm{ex}}$ stands for the angular magnetic moment when the system enters the exchange dominated regime. For example, in Mn$_2$Ru$_x$Ga, since Mn spins are assumed to demagnetize at the same rate, this condition reduces to $S^{\rm{ex}}_a+S_b^{\rm{ex}} \approx (S_a(0)+S_b(0))\exp(-t/\tau_a)<0$. Namely, only for a starting temperature below the compensation temperature, $S_a(0)+S_b(0)<0$, conditions for switching are fulfilled, in complete agreement to experimental observations \cite{Banerjee2020}. Yet, the exchange relaxation regime needs to be active for a significant time in order for the magnetization to switch. We compare the time scales associated to both the relativistic and exchange relaxation. Relativistic relaxation rate follows $\Gamma_a^r = 2 \lambda_a k_B (T-T_c^a) m^z_a$, which is strongly reduced by the ultrafast dynamics of $m^z_a$, in only a few hundred of femtoseconds $\Gamma_a^r\rightarrow 0$. Whereas $\Gamma^{\rm{ex}}$ rather follows the dynamics of the temperature $T$, and therefore decays slower than $\Gamma_a^r$ (Fig. \eqref{fig:ASD-model-simulations}(c) and (d)).  

The characteristic time scale of the electron and lattice temperature is described by the TTM and for common parameter values in the range of 2 $-$ 3 ps. As the temperature reduces, the exchange relaxation drives the system towards $(T-T_c^a) m^z_a = (T-T_c^b) m^z_b$. For $T_c^b < T < T_c^a$, the exchange relaxation drives the system back into an antiferromagnetic, but switched sate. As the magnetization builds up in the opposite direction, $\Gamma^{\rm{ex}}$, decreases and the relativistic relaxation takes over. Interestingly, our theory predicts that for systems where $T_c^a \approx T_c^b$ switching would be unlikely, which has been recently demonstrated in rare-earth free synthetic ferrimagnets \cite{Liao2019}. Further, one can accelerate the transition from exchange relaxation dominated to the relativistic regime, and speed up complete switching by increasing difference between $T_c^a$ and $T_c^b$, an effect which has been observed by the substitution of Fe by Co, namely, GdFe by GdCo \cite{Ceballos2021}. While GdFe recovers in tens of picoseconds, GdCo alloys only need of a couple of picoseconds. 
     
To summarize, in this work we have proposed a general macroscopic theory for the magnetization dynamics of ferrimagnetic materials driven by femtosecond laser photo-excitation. Our theory reproduces quantitatively all stages of the switching process observed in experiments. Notably, we have directly compared our theory to computer simulations using atomistic spin dynamics methods for both GdFeCo and Mn$_2$Ru$_x$Ga alloys. The magnetization dynamics transits from a relativistic relaxation path to an exchange dominated regime due to the strong enhancement of the exchange relaxation. We demonstrate that switching occurs when the sublattice magnetization reaches a threshold value. These criteria substitute previous ones and pave the way for the discovery of new class of ferrimagnets showing switching.

\emph{Acknowledgement.} The authors thank Ilie Radu and Jon Gorchon for useful discussions and critical reading of the manuscript. We gratefully acknowledge support by the Deutsche Forschungsgemeinschaft through SFB/TRR 227  "Ultrafast Spin Dynamics", Project A08.

\bibliography{cleanLIB}

\end{document}